\documentclass[12pt]{iopart}

\def\Journal#1#2#3#4#5{#5 {\it #1} #2 {\bf #3} #4}
\def\JP{J. Phys}
\def\PL{Phys. Lett}
\def\PR{Phys. Rev.}
\def\PRL{Phys. Rev. Lett.}
\newcommand{\ber}{\begin{eqnarray}}
\newcommand{\eer}{\end{eqnarray}}
\newcommand{\vpaa}{{\bf p_\alpha}}
\newcommand{\vpbb}{{\bf p_\beta}}
\newcommand{\vp}{{\bf p}}
\newcommand{\vq}{{\bf q}}
\newcommand{\vpq}{{\bf p_q}}
\newcommand{\vpm}{{\bf p_M}}
\newcommand{\vpb}{{\bf p_B}}
\newcommand{\ptc}{{p_{\perp 0}}}
\newcommand{\ptcp}{{p^\prime_{\perp 0}}}
\newcommand{\ptcc}{{p_{\perp 0,c}}}

\newcommand{\pcq}{{p^{coal.}_q({\bf p_q})}}

\newcommand{\nqm}{{N_{q_M}}}
\newcommand{\nqb}{{N_{q_B}}}
\newcommand{\nqmc}{{N^3_{q_M}}}
\newcommand{\nqbc}{{N^3_{q_B}}}
\newcommand{\pp}{p_\perp}
\newcommand{\vtwoq}{{v_{2,q}}}
\newcommand{\vtwoc}{{v_{2,c}}}
\newcommand{\vpc}{{\bf p_c}}
\newcommand{\vpd}{{\bf p_D}}
\newcommand{\vpj}{{\bf p_\psi}}
\newcommand{\pcc}{{p^{coal.}_c({\bf p_c})}}
\newcommand{\zqd}{z_{q_D}}
\newcommand{\zcd}{z_{c_D}}
\newcommand{\zqdc}{{z^3_{q_D}}}
\newcommand{\zcdc}{{z^3_{c_D}}}

\begin{document}

\title{Quark Coalescence with Quark Number Conservation 
and the Effect on Quark-Hadron Scaling}

\author{Zi-Wei Lin}

\address{Department of Physics, East Carolina University, 
C-209 Howell Science Complex, Greenville, NC 27858}
\ead{linz@ecu.edu}
\begin{abstract}
We develop a new formulation of the quark coalescence model by
including the quark number conservation in order to describe the
hadronization of the bulk of the quark-gluon plasma.  The scalings
between hadron and quark phase space distributions are shown to depend
on the transverse momentum.  For hard quarks, our general scalings
reproduce the usual quadratic scaling relation for mesons and the
cubic scaling relation for baryons.  
For softer quarks, however, the inclusion of the quark
number conservation leads to a linear scaling for the hadron species
that dominates the quark number of each flavor, while the scalings of
non-dominant hadrons depend on the coalescence dynamics.  For charm
mesons, we find that the distribution of soft $D$ mesons does not
depend on the light quark distribution but the distribution of soft
$J/\psi$ mesons is inversely correlated to the light quark
distribution.
\end{abstract}

\pacs{12.38.Mh, 25.75.Ld, 25.75.Nq}


\section{Introduction}

For the dense matter created in ultra-relativistic heavy ion collisions such 
as those at the Relativistic Heavy Ion Collider (RHIC) and the Large Hadron 
Collider (LHC), the quark coalescence model provides a hadronization model.  
When the quark-gluon plasma reaches the hadronization hypersurface after 
expansion,  a quark and an anti-quark can form a meson 
while three quarks can form a baryon. 
As a result, meson and baryon momentum distributions are respectively
proportional to the product of two and three quark momentum
distributions. This has led to the scaling of hadron elliptic flows 
with the valence quark number in the hadron 
\cite{Molnar:2003ff,Adler:2003kt,Abelev:2007rw}, 
a novel scaling of heavy hadrons when the coalescing quarks have 
different masses \cite{Lin:2003jy}, and an enhanced baryon-to-meson
ratio \cite{Fries:2003vb,Greco:2003xt,Adler:2003cb,Adams:2006ke}. 

However, the current formulation of the quark coalescence model does not 
conserve the quark number of each flavor, i.e., it violates unitarity. 
For example, suppose there are an equal number of quarks and anti-quarks and 
they all coalesce to mesons, the total number of mesons after coalescence 
will then be equal to the initial number of quarks,  
and doubling the initial number of quarks and anti-quarks 
will lead to twice as many mesons. 
However, in the current formulation of the quark coalescence model,
doubling the initial distribution function of quarks and anti-quarks 
will lead to four times as many mesons. 
Therefore, the current coalescence formulation only applies 
to partons above a moderate transverse momentum scale $\ptc$,  
where the coalescence probability is small and 
the effect from unitarity violation is expected to be unimportant. 

For the bulk of partons in the quark-gluon
plasma that are soft with transverse momenta well below the scale
$\ptc$, we may expect quark coalescence to be the dominant hadronization 
process because, due to the much higher density of soft partons, 
it is much easier to find two or three soft quark or anti-quark partners 
to form a hadron. However, a formulation including the quark number 
conservation is necessary to address the coalescence of soft quarks. 
Note that quarks or anti-quarks that cannot find coalescence partner(s) would 
hadronize via other processes such as independent fragmentation. 
Several earlier attempts to include unitarity have been made. 
In the algebraic coalescence rehadronization (ALCOR) model, the 
quadratic and cubic scaling relations are used to describe 
meson and baryon multiplicities, respectively; and quark number
conservations are enforced by the normalization constants \cite{Zimanyi:ky}. 
Another earlier study to include unitarity \cite{Yang:2005dg}
addressed the total multiplicities of different hadrons from quark coalescence
and found that, as expected, the total hadron multiplicity is linearly
proportional to the initial quark number.
In the Monte Carlo transport approach, a multi-phase transport (AMPT)
model \cite{Lin:2004en} converts nearby quarks into hadrons one-by-one
and therefore automatically satisfies the quark number conservation;
however, the model needs to be improved in its quark coalescence
criterion that determines the average parton density at coalescence
and the effective equation of state in the model \cite{Zhang:2008zzk}.

In this study we incorporate the quark number conservation 
in the quark coalescence model so that 
the model can be applied to quarks at all transverse momenta. 
The general results as functions of quark transverse momentum are 
expected to yield the familiar quadratic and cubic scaling behavior
\cite{Molnar:2003ff,Lin:2003jy,Fries:2003vb,Greco:2003xt} for hard
quarks. On the other hand, for soft quarks where the coalescence
probability approaches one, the results should conserve the quark
numbers of each flavor. 

\section{The formulation including the quark number conservation}

Let us write 
$f(\vp,x)=(2\pi)^3 dN/(d^3x d^3p)$
as the phase space distribution of particles. 
A convenient formula for the coalescence production of 
meson $M$ via $\alpha \beta \to M$ is \cite{Dover:1991zn}
\ber
E \frac{dN_M(\vp)}{d^3p} = 
\int  \frac{d\sigma^\mu p_\mu}{(2\pi)^3} 
\int  d^3  q g_M   \left|\Psi_\vp \, (\vq)\right|^2 
f_\alpha(\vpaa,x) f_\beta (\vpbb,x) ,\label{dnm}
\eer
where $\vp \equiv \vpaa+\vpbb$, $\vq \equiv \vpaa-\vpbb$, 
and $g_M$ is the statistical factor for forming the meson. 
The meson wave-function is normalized as 
$\int  d^3q |\Psi_\vp(\vq)|^2=1$, and the first integration 
in Eq.~(\ref{dnm}) runs over a 3-dimensional space-time hypersurface. 
Assuming that the hadronization time scale is short compared to the
time scale for the expansion of the system when coalescence takes place, 
so that the parton hypersurface during coalescence is unchanged and 
the hypersurface of formed hadrons is the same as 
that of the coalescing quarks, we can then write $E~dN_M(\vp)/d^3p= 
\int  d\sigma^\mu p_\mu f_M(\vp,x)/(2\pi)^3$,
where the integration is over the same hypersurface. 
The coalescence yield from a local hypersurface can then be written as
\ber
f_M (\vp,x) = 
\int  d^3  q \; g_M \left|\Psi_\vp \, (\vq)\right|^2 
f_\alpha(\vpaa,x) f_\beta (\vpbb,x).
\eer

To include the quark number conservation, 
we need to take into account the fact that 
the quark number decreases as quarks coalescence into hadrons. 
We thus consider the time evolution of the coalescence process.
To differentiate from the time variable used to represent 
the hypersurface, let us use $t_c$ to represent 
the time variable of the coalescence process.   
For a local hypersurface at $x$, we write
\ber
f^\prime_M(\vp,x,t_c) = 
\int  d^3  q  C_M(\vp,\vq,x,t_c )
f_\alpha(\vpaa,x,t_c ) f_\beta (\vpbb,x,t_c ),
\label{coalm1}
\eer
where $f^\prime_M(\vp,x,t_c)  \equiv  df_M(\vp,x,t_c)/d t_c$, 
and the phase space distributions such as 
$f_\alpha(\vpaa,x,t_c)$ and $f_M(\vp,x,t_c)$ are time-dependent.
The term $C_M(\vp,\vq,x,t_c)$ represents the probability for the coalescence 
production of meson $M$ that could depend on $x$ and time $t_c$.

When the internal momenta of the coalescing quarks in a hadron can be 
neglected, we can write
\ber
C_M(\vp,\vq,x,t_c)=c_M(\vp,x,t_c) \delta(\vq)
\eer
for meson $M$. Note that the effect of the internal momenta of coalescing 
quarks may be important for soft hadrons \cite{Lin:2003jy,Greco:2004ex};
however it is neglected in this study so that 
we can obtain analytical results.

When valence quarks have the same mass, 
Eq.~(\ref{coalm1}) reduces to 
\ber
f^\prime_M(\vpm,x,t_c)= 
c_M(\vpm,x,t_c) f_\alpha(\vpq,x,t_c)f_\beta(\vpq,x,t_c),
\label{dfmdt}
\eer
where $\vpm=\nqm \vpq$ with $\nqm=2$, 
and $c_M(\vpm,x,t_c)$ represents the coalescence coefficient for the meson.
Similarly, for baryon productions via $\alpha+\beta+\gamma \to B$,
we can write
\ber
f^\prime_B(\vpb,x,t_c)=c_B(\vpb,x,t_c)  
f_\alpha(\vpq,x,t_c)  
f_\beta(\vpq,x,t_c) f_\gamma(\vpq,x,t_c) ,
\label{dfbdt}
\eer
where $\vpb=\nqb \vpq$ with $\nqb=3$,
and $c_B(\vpb,x,t_c)$ represents the coalescence coefficients for the baryon.

The local conservations of quark numbers of each flavor during the coalescence 
process are given by a set of equations. 
As an example, when a meson $M$ contains one constituent 
quark of flavor $\alpha$ and a baryon $B$ consists of three 
constituent quarks of different flavors $\alpha$, $\beta$ and $\gamma$, 
the local number conservation for quark flavor $\alpha$ in the quark-hadron 
system can be expressed as 
\ber
dN_\alpha(\vpq,x,t_c)= -dN_M(\vpm,x,t_c)-dN_B(\vpb,x,t_c). 
\eer
Since $dN_M \propto f_M d^3 p_M = \nqmc f_M d^3 p_q$
and $dN_B \propto f_B d^3 p_B = \nqbc f_B d^3p_q$,
the above equation leads to
\ber
f^\prime_\alpha(\vpq,x,t_c)= 
-\nqmc f^\prime_M(\vpm,x,t_c)-\nqbc f^\prime_B(\vpb,x,t_c). 
\label{cons}
\eer
Note that the third powers on $\nqm$ and $\nqb$ in the above equation result
from the three-dimension nature of momentum, but they are unrelated to 
the number of constituent quarks in the hadron.

For a local coalescence process that starts at $t_0$ and 
ends at $t_F$, the initial conditions for coalescence are given by
\ber
f_\alpha (\vpq,x,t_0) \equiv f_0 (\vpq,x), 
f_M (\vpm,x,t_0)=0, f_B (\vpb,x,t_0)=0
\label{init}
\eer
for quark flavor $\alpha$, meson $M$ and baryon $B$, respectively.  
In the following we mostly use the simplified notations, where
the label $x$ is omitted in all functions and 
the variable $t_c$ is rewritten as $t$.

\section{Consideration of one meson and one baryon species}

We consider the coalescence productions of one meson species 
and one baryon species from one quark flavor via $q+\bar q \to M$, 
$q+q+q \to B$, and $\bar q+\bar q+\bar q \to \bar B$. 
Let us assume zero baryon chemical potential for simplicity. 
For the quark distribution 
at momentum $\vpq$, we have the following rate equations
\ber
f^\prime_M(\vpm,t)&=& c_M(\vpm,t) f^2_q(\vpq,t), \nonumber \\
f^\prime_B(\vpb,t)&=& c_B(\vpb,t) f^3_q(\vpq,t).
\label{eqn1}
\eer

The conservation of quark numbers is given by
\ber
f^\prime_q(\vpq,t)= 
-\nqmc f^\prime_M(\vpm,t)-3 \nqbc f^\prime_B(\vpb,t). 
\label{eqn2}
\eer
Therefore the quark distribution is given by
\ber
f^\prime_q (\vpq,t)=  
- \nqmc  c_M(\vpm,t) f^2_q(\vpq,t)  
- 3\nqbc  c_B(\vpb,t) f^3_q(\vpq,t). 
\label{eqn3}
\eer
Note that Eq.~(\ref{eqn2}) is different from Eq.~(\ref{cons}) because 
a baryon $B$ in the case of Eq.~(\ref{eqn2}) contains three 
constituent quarks of the same flavor, 
while a baryon in the case of Eq.~(\ref{cons})
contains only one constituent quark of flavor $\alpha$.

\subsection{Solutions when coalescence coefficients 
have the same time-dependence}
\label{sec31}

If the meson and baryon coalescence coefficients 
have the same time-dependence, we can define
\ber
r(\vpq) \equiv \frac {3\nqbc c_B(\vpb, t)}{\nqmc c_M(\vpm, t)}.
\label{ratio}
\eer 
Eq.~(\ref{eqn3}) then becomes 
\ber
\frac {f^\prime_q (\vpq,t)}{f^2_q(\vpq,t) + r(\vpq) f^3_q(\vpq,t)}
=  - \nqmc  c_M(\vpm,t),
\eer 
which can be integrated to yield the following solution for $f_q(\vpq,t)$:
\ber
& &  r(\vpq) \ln  
\left [ \left ( \frac{ 1+r(\vpq) f_0 (\vpq)} { 1+r(\vpq) f_q(\vpq,t)} \right ) \left ( \frac{f_q (\vpq,t)} {f_0 (\vpq)} \right ) \right ] \nonumber \\
&+& \frac{1}{f_q (\vpq,t)} -\frac{1}{f_0 (\vpq)}
=I_M(\vpm,t), \nonumber \\
& {\rm with~} & I_M(\vpm,t) \equiv \nqmc  
\int_{t_0}^t c_M(\vpm,u) du.
\label{solution}
\eer
We see that above factor $I_M(\vpm,t)$ is proportional to the time integral of 
the coalescence coefficient for the meson. For hard quarks, we shall see that 
the final value of this factor is given by $I_M(\vpm,t_F) \simeq \nqmc g_M$ 
according to Eq.~(\ref{normcm}).

\subsubsection{The limit that mesons dominate}
\label{sec311}

We first consider the limit $r(\vpq) \to 0$, where meson productions 
dominate. In this limit, the left-hand side of the solution in 
Eq.~(\ref{solution}) reduces to the form 
$1/f_q-1/f_0+r\ln(f_q/f_0)+{\cal O}(r^2)$, 
and we can obtain the solutions at the leading order in $r(\vpq)$ as
\ber
f_q (\vpq,t) &\simeq & 
\frac {f_0(\vpq)}{1+f_0(\vpq) I_M(\vpm,t)}, 
\label{fq1} \\
f_M (\vpm,t) &\simeq &  \frac {f^2_0 (\vpq) I_M(\vpm,t)}
{\nqmc \left [ 1+f_0 (\vpq) I_M(\vpm,t) \right ] }, \label{fm1} \\
f_B (\vpb,t) &\simeq & \frac {f^2_0 (\vpq) r(\vpq)} {6\nqbc}  \left \{  1-\frac{1} {\left [ 1+f_0 (\vpq ) I_M(\vpm,t) \right ] ^2}  \right \} .
\label{fb1}
\eer

Therefore the coalescence probability of quarks at momentum $\vpq$ is 
given by
\ber
\pcq = 1- \frac{f_q(\vpq,t_F)}{f_0 (\vpq)}
= 1-\frac {1}{1 + f_0 (\vpq) I_M(\vpm,t_F)}.
\label{pcoal}
\eer
Since we expect quarks well above the scale $\ptc$ to have a very small 
coalescence probability and quarks well below the scale $\ptc$ to have 
a coalescence probability of almost one, 
we may choose to define the scale $\ptc$ so that it gives $\pcq =1/2$;
$\ptc$ then corresponds to the quark transverse momentum 
where $f_0 (\vpq)I_M(\vpm,t_F)=1$.

For hard quarks, those with transverse momenta well above the scale
$\ptc$, the small value of $f_0 (\vpq)$ at high $\pp$ leads to 
$f_0 (\vpq) I_M(\vpm,t_F) \ll  1$, 
Eq.~(\ref{pcoal}) then gives  
\ber
p^{coal.}_{hard~q}({\bf p_q}) \simeq f_0 (\vpq) I_M(\vpm,t_F) \ll  1.
\eer
Note that the coalescence probability of hard quarks is proportional to its 
distribution function $f_0 (\vpq)$, and this is why the hard meson and baryon 
elliptic flows are enhanced from the quark elliptic flow 
by a factor of 2 and 3 respectively, as seen in Eq.~(\ref{v2hard}).
Also note that, given the small coalescence probability, 
hard quarks mostly hadronize via other processes such as fragmentation. 

We can obtain the final hadron distributions 
from the coalescence of hard quarks as 
\ber
f^{hard}_M (\vpm,t_F)  &\simeq & 
f^2_0 (\vpq) I_M(\vpm,t_F)/\nqmc, \nonumber \\
f^{hard}_B (\vpb,t_F)  &\simeq & 
f^3_0 (\vpq) I_B(\vpb,t_F)/(3\nqbc ), \nonumber \\
{\rm with~} 
I_B(\vpb,t)  & \equiv &  3\nqbc  \int_{t_0}^t c_B(\vpb,u) du.
\label{fmb2}
\eer
Note that for hard quarks we have 
$f_q (\vpq,t) \simeq  f_0 (\vpq)$, 
the time-integration of Eq.~(\ref{coalm1}) then gives 
\ber 
\int_{t_0}^{t_F} C_M(\vp,\vq,x,t_c) dt_c \simeq g_M |\Psi_\vp(\vq)|^2.   
\eer
When the internal momenta of the coalescing quarks in a hadron are neglected,  
the above reduces to
\ber 
\int_{t_0}^{t_F} c_M(\vpm,t) dt \simeq g_M.
\label{normcm}
\eer
Similarly, we also have 
\ber
\int_{t_0}^{t_F} c_B(\vpb,t) dt \simeq g_B
\label{normcb}
\eer
for hard quarks. The above two relations may serve as normalization
relations for the coalescence coefficients for mesons and baryons, and
they also help to determine the coalescence times $t_0$ and $t_F$
if the coalescence coefficients are known.
As a result, the functions $I_M(\vpm,t_F)$, $I_B(\vpb,t_F)$ 
and $r(\vpq)$ for hard quarks are constants that do not depend on $\phi$, 
the azimuthal angle of the momentum vector $\vpq$ in the transverse plane of 
a heavy ion collision. We then have
\ber
f^{hard}_M (\vpm,t_F) & \propto & f^2_0 (\vpq), \nonumber \\
f^{hard}_B (\vpb,t_F) & \propto & f^3_0 (\vpq).
\eer
These are just the scaling relations of the previous quark coalescence model 
\cite{Molnar:2003ff,Lin:2003jy,Fries:2003vb,Greco:2003xt}.
Consequently, when the quark momentum distribution has 
an azimuthal asymmetry that is dominated by the $\cos (2\phi)$ term, 
the above scalings lead to the following quark-number scaling 
of the hadron elliptic flows ($v_2$) 
in ultra-relativistic heavy ions collisions \cite{Molnar:2003ff}:
\ber
v^{hard}_{2,M} (\pp )  & \simeq &  2 \vtwoq (\pp  /  \nqm ), 
\nonumber \\
v^{hard}_{2,B} (\pp )  & \simeq &  3 \vtwoq (\pp  /  \nqb ). 
\label{v2hard}
\eer

For soft quarks, those with transverse momenta well below the scale $\ptc$, 
they mostly coalesce and thus $\pcq \simeq  1$, provided that 
$f_0 (\vpq) I_M(\vpm,t_F)  \gg  1$. 
This is consistent with the fact that $f_0 (\vpq)$ increases rapidly 
as the transverse momentum gets lower.
The final hadron distributions from the coalescence of soft quarks 
are then given by 
\ber
f^{soft}_M (\vpm,t_F) &\simeq & 
f_0 (\vpq)/\nqmc, \nonumber \\
f^{soft}_B (\vpb,t_F) &\simeq & f^2_0 (\vpq) r(\vpq)/(6\nqbc).
\label{fb4}
\eer
The above linear scaling for the dominant mesons 
is the result of including the quark number conservation in the formulation. 
Integrating the above meson distribution over the meson three-momentum yields 
the quark number conservation relation $N_M \simeq N_{q,0}$, 
where $N_{q,0}$ represents the total number of quarks 
just before coalescence.

When the quark momentum distribution has an azimuthal asymmetry that is 
dominated by the $\cos (2\phi)$ term, the above relationships in 
Eqs.~(\ref{fb4}) between the quark and hadron momentum distributions 
then lead to the following scaling between the 
meson elliptic flows and those of the constituent quarks:
\ber
v^{soft}_{2,M} (\pp)  \simeq  \vtwoq (\pp  /  \nqm  ).
\label{v2msoft}
\eer
On the other hand, the scaling of baryons, which are the non-dominant
hadrons in this case, is quadratic. 
If $r(\vpq)$ for soft quarks also does not depend on the angle $\phi$, 
we have
\ber 
v^{soft}_{2,B} (\pp)  \simeq  2 \vtwoq (\pp  /  \nqb  ). 
\label{v2bsoft}
\eer

Comparing the scalings of Eqs.(\ref{v2msoft}-\ref{v2bsoft}) 
with those in Eq.(\ref{v2hard}), 
we find that the scalings of hadron elliptic flows with the quark elliptic flow
for soft quarks are weaker than those for hard quarks. 
For example, Eq.~(\ref{v2msoft}) gives 
$v^{soft}_{2,M} (\pp) \simeq \vtwoq (\pp/2)$ for soft mesons, while 
the scaling for hard mesons, $v^{hard}_{2,M} (\pp) \simeq 2 \vtwoq (\pp/2)$,
is twice as strong. 
The weaker scaling can be understood from Eq.~(\ref{fq1}), 
where the term $f_0 (\vpq)$ in the denominator means that 
the azimuthal asymmetry in the quark momentum distribution 
during the coalescence process will be reduced from its initial magnitude 
(provided that the coalescing coefficient in the term $I_M(\vpm,t)$  
in the denominator does not depend on the azimuthal angle). 
For hard quarks, this reduction is negligible because the condition
$f_0 (\vpq) I_M(\vpm,t)  \ll  1$ is true throughout  
the coalescence process. For soft quarks, however, the azimuthal 
asymmetry in the quark momentum distribution gradually vanishes during 
coalescence, leading to weaker scalings of soft hadron elliptic flows with 
the quark elliptic flow. 

\subsubsection{The limit that baryons dominate}

Let us now consider the opposite limit 
$r(\vpq)  \to  \infty$ 
by keeping $c_B(\vpb,t)$ fixed while decreasing $c_M(\vpm,t)$.
In this case, baryon productions dominate, 
and we obtain the solutions at the leading order in $1/r(\vpq)$ as
\ber
f_q (\vpq,t) &\simeq & 
\frac {f_0 (\vpq)}{\sqrt {1+2f^2_0 (\vpq) I_B(\vpb,t)}}, 
\nonumber \\
f_M (\vpm,t) &\simeq &  \frac 
{\ln  \left [ 1+2f^2_0 (\vpq) I_B(\vpb,t) \right ] }
{2 \nqmc r(\vpq)}, 
\nonumber \\
f_B (\vpb,t) &\simeq & \frac {f_0 (\vpq)} 
{3 \nqbc} 
\left ( 1-\frac{1} {\sqrt {1+ 2f^2_0 (\vpq ) I_B(\vpb,t)}}  \right ) . 
\label{fqmb2}
\eer

The scale $\ptc$ in this case corresponds to the quark transverse momentum 
where $f^2_0 (\vpq) I_B(\vpb,t_F)=3/2$.
For hard quarks well above the transverse momentum $\ptc$, 
the final hadron distributions are the same as Eq.~(\ref{fmb2}).
However, hadron distributions from the coalescence of soft quarks are given by
\ber
f^{soft}_M (\vpm,t_F) &\simeq & 
\ln  \left [ 2 f^2_0 (\vpq) I_B(\vpb,t_F) \right ]
/\left [2 \nqmc r(\vpq) \right ], \nonumber \\
f^{soft}_B (\vpb,t_F) &\simeq & f_0 (\vpq)/(3\nqbc).
\label{fqmb3}
\eer
We again see a linear scaling due to the quark number conservation, 
but in this case it is for the dominant baryons;  
and integrating the above baryon distribution over the baryon three-momentum 
yields the quark number conservation relation $N_B \simeq N_{q,0}/3$. 
Note that the scaling of the non-dominant mesons with the quark distribution 
in this case is neither linear nor quadratic.  

Similar to the $r(\vpq) \to 0$ case, 
from Eq.~(\ref{fqmb2}) we also see that the azimuthal asymmetry in the
momentum distribution of soft quarks 
gradually vanishes during the coalescence process, 
provided that the coalescing coefficient in the term $I_B(\vpb,t)$  
in the denominator does not depend on the azimuthal angle. This leads to 
weaker scalings of hadron elliptic flows with the quark elliptic flow for 
soft quarks than for hard quarks. For example, Eq.~(\ref{fqmb3}) gives 
$v^{soft}_{2,B} (\pp) \simeq \vtwoq (\pp/3)$ for soft baryons, while 
the scaling for hard baryons, $v^{hard}_{2,B} (\pp) \simeq 3 \vtwoq (\pp/3)$,
is three times as strong. 

\subsection{Solutions for arbitrary coalescence coefficients in the
  limit that mesons dominate}
\label{sec32}

Without assuming that the baryon and meson coalescence coefficients  
have the same time-dependence, we now consider 
Eqs.~(\ref{eqn1}-\ref{eqn3}) in the limit that mesons dominate. 
In this case, the leading-order solutions of $f_q (\vpq,t)$ and 
$f_M (\vpm,t)$ are still given by Eq.~(\ref{fq1}) 
and Eq.~(\ref{fm1}), respectively. However, the baryon distribution is given by
\ber
f_B (\vpb,t) \simeq   
\int_{t_0}^t  \frac {f^3_0 (\vpq) c_B(\vpb,u) du}
{\left [1+f_0(\vpq) I_M(\vpm,u)\right ]^3}.
\label{fb10}
\eer

For hard quarks well above the scale $\ptc$, 
the above is the same as the baryon distribution in Eq.~(\ref{fmb2}). 
For soft quarks, however, the scaling behavior of the non-dominant baryons 
depends on details of the coalescence dynamics, which are represented 
by the coalescence coefficients $c_M(\vpm,t)$ and $c_B(\vpb,t)$. 

For example, 
if $c_B(\vpb,t) =  \epsilon~ c_M(\vpm,t) I_M(\vpm,t)$ 
with $\epsilon \ll 1$, we have
\ber
f_B (\vpb,t) \simeq \frac {\epsilon f^3_0 (\vpq) I^2_M(\vpm,t)} 
{2 \nqmc \left [1+f_0 (\vpq ) I_M(\vpm,t) \right ]^2}  .
\eer
For soft quarks, this leads to 
\ber
f^{soft}_B (\vpb,t_F)  & \simeq &  
\epsilon f_0 (\vpq)/(2 \nqmc) , \nonumber \\
v^{soft}_{2,B} (\pp)  & \simeq & \vtwoq (\pp/\nqb) . 
\eer
Note that for soft mesons, which dominate the quark numbers, their
distribution is the same as that in Sec.~\ref{sec311} and their
elliptic flow scales the same way as in Eq.~(\ref{v2msoft}), i.e.,
\ber
v^{soft}_{2,M} (\pp)  \simeq  \vtwoq (\pp  /  \nqm  ).
\eer

On the other hand, if $c_B(\vpb,t)  \propto  c_M(\vpm,t)$ as assumed 
in Sec.~\ref{sec31},
the baryon distributions of Eq.~(\ref{fb1}) and Eq.~(\ref{fb4}) 
are reproduced, where soft baryons follow a different 
scaling as given by Eq.~(\ref{v2bsoft}).
Therefore the dynamics of coalescence can be probed by studying hadrons 
from the coalescence of quarks below the scale $\ptc$. 

\section{Charm mesons}

We consider the coalescence production of charm mesons 
via $\bar q+ c \to D$, $q+\bar c \to \bar D$, and $c+\bar c \to J/\psi$, 
in addition to the coalescence production of a light meson species 
via $q+\bar q \to M$. 
Assuming zero baryon chemical potential for simplicity, 
we solve the following rate equations:
\ber
f^\prime_M(\vpm,t)&=& c_M(\vpm,t) f^2_q(\vpq,t), \nonumber \\
f^\prime_D(\vpd,t)&=& c_D(\vpd,t) f_q(\vpq,t) f_c(\vpc,t), 
\nonumber \\
f^\prime_\psi(\vpj,t)&=& c_\psi(\vpj,t) f^2_c(\vpc,t),  
\eer
together with the following equations that represent 
the conservations of light quark and charm quark numbers:
\ber
f^\prime_q(\vpq,t)&=& 
-\nqmc f^\prime_M(\vpm,t)-f^\prime_D(\vpd,t)/\zqdc , \nonumber \\ 
f^\prime_c(\vpc,t)&=& 
-f^\prime_D(\vpd,t)/\zcdc-\nqmc f^\prime_\psi(\vpj,t). 
\label{consc}
\eer
Note that the above Eq.~(\ref{consc}) results from the conservation relations 
$dN_q(\vpq)=-dN_M(\vpm)-dN_D(\vpd)$ 
and $dN_c(\vpc)=-dN_D(\vpd)-dN_\psi(\vpj)$. In the above, 
$\zqd=m_q/(m_c+m_q)$, $\zcd=m_c/(m_c+m_q)$, where $m_q$ and $m_c$ are 
the effective masses of the valence light quark and charm quark, respectively. 
The light quark momentum involved in the above equations 
is $\vpq$, and as a result the charm quark momentum in consideration is given 
by $\vpc=\vpq m_c/m_q$, the $D$ meson momentum is $\vpd=\vpc/\zcd$, and 
the $J/\psi$ meson momentum is $\vpj=\nqm \vpc$ \cite{Lin:2003jy}. 
This way the light quark momentum in a $D$ meson is $\vpq=\zqd \vpd$ 
and the charm quark momentum in a $D$ meson is $\vpc=\zcd \vpd$, 
so that they have the same velocity before coalescing into a charm meson 
\cite{Lin:2003jy}. 

Assuming that light mesons dominate over $D$ mesons in the light quark sector 
and $D$ mesons dominate over $J/\psi$ mesons in the charm quark sector, 
we obtain the following solutions at the leading order in 
$f_{c0}(\vpc)/f_0(\vpq)$: 
\ber
f_c (\vpc,t) \simeq f_{c0} (\vpc)
e^ {-\int_{t_0}^t c_D(\vpd,u) f_q (\vpq,u)/\zcdc du},
\eer
where $f_{c0} (\vpc) \equiv  f_c (\vpc,t_0)$ represents the  
charm quark distribution just before coalescence, 
and $f_q (\vpq,t)$ is given by Eq.~(\ref{fq1}) at the leading order.
The light meson distribution $f_M (\vpm,t)$ 
is given by Eq.~(\ref{fm1}), and the $D$ meson distribution is given by 
\ber
f_D (\vpd,t) \simeq  f_{c0} (\vpc) \zcdc 
\left (1-e^ {-\int_{t_0}^t c_D(\vpd,u)f_q (\vpq,u)/\zcdc du}  \right ) . 
\eer

\subsection{Solutions when coalescence coefficients 
have the same time-dependence}

We now consider the case where the coalescence coefficients  
have the same time-dependence, and define 
\ber
r_D(\vpd)  & \equiv &  
\frac {c_D(\vpd, t)}{c_M(\vpm, t) \zcdc  \nqmc} , 
\nonumber \\
r_\psi(\vpj)  & \equiv &  
\frac {c_\psi(\vpj, t)\zcdc \nqmc}{c_D(\vpd, t)} .
\eer 
The solutions at the leading order in $f_{c0}(\vpc)/f_0(\vpq)$ then simplify to
\ber
f_c (\vpc,t) &\simeq & \frac {f_{c0} (\vpc)}
{\left [ 1+f_0 (\vpq) I_M(\vpm,t) \right ]^{r_D(\vpd)}}, 
\label{fc} \\
f_D (\vpd,t) &\simeq & f_{c0} (\vpc) \zcdc
\left \{ 1- \frac{1}{ \left [ 1+f_0 (\vpq) I_M(\vpm,t) \right ]^{r_D(\vpd)}} \right \}, \nonumber \\
f_\psi (\vpj,t) &\simeq & 
\frac {f^2_{c0} (\vpc) r_D(\vpd) r_\psi(\vpj)}
{f_0 (\vpq) \left [1-2r_D(\vpd) \right ] \nqmc } \nonumber \\
&\times &
\left \{ \left [1+f_0 (\vpq) I_M(\vpm,t) \right ]^{1-2r_D(\vpd)}-1 \right \}.
\label{fd}
\eer

Therefore the coalescence probability of charm quarks at momentum $\vpc$ is
\ber
\pcc = 1- \frac{f_c(\vpc,t_F)}{f_0 (\vpc)}
= 1-\frac {1}{\left [ 1+f_0 (\vpq) I_M(\vpm,t_F) \right ]^{r_D(\vpd)}}.
\label{ccoal}
\eer

If we define the scale $\ptcc$ that separates the soft and hard charm
quarks according to $\pcc =1/2$, this scale $\ptcc$ is then given
by $\ptcp m_c/m_q$, where the scale $\ptcp$ is the light quark
transverse momentum that corresponds to 
\ber
f_0 (\vpq) I_M(\vpm,t_F)=2^{1/r_D(\vpd)}-1.
\eer 
Note that the coalescence probability of light quarks is still given 
by Eq.~(\ref{pcoal}), therefore the scale $\ptc$ that separates the
soft and hard light quarks, which corresponds to 
$f_0 (\vpq) I_M(\vpm,t_F)=1$, is different from the scale $\ptcp$. 
Also, the scale for charm quarks, $\ptcc=\ptcp m_c/m_q$, 
could be a sizeable momentum because of the large quark mass $m_c$, 
therefore the transverse momentum range for {\it soft} charm hadrons 
could be sizeable, and it could be easier to observe the 
scaling relations of soft mesons in heavy flavor observables. 

For the coalescence of hard quarks, we have
\ber
f^{hard}_D (\vpd,t_F) &\simeq & 
f_0 (\vpq) f_{c0} (\vpc) \zcdc r_D(\vpd) I_M(\vpm,t_F), 
\nonumber \\
f^{hard}_\psi (\vpj,t_F) &\simeq & 
f^2_{c0} (\vpc) r_\psi(\vpj) r_D(\vpd) I_M(\vpm,t_F)
/\nqmc ,
\eer
which reproduce the previous scaling relations \cite{Lin:2003jy}.

For soft quarks, we first note that, according to Eq.~(\ref{fc}) 
and the solution of $f_q (\vpq,t)$ in Eq.~(\ref{fq1}), 
the assumptions that we made 
in obtaining the solutions of Eqs.~(\ref{fc}-\ref{fd})
are valid when $r_D(\vpd) \geq 1$. 
If $r_D(\vpd) < 1$, the charm quark distribution decreases with time 
at a slower rate than the light quark distribution, and therefore 
the number of charm quarks would be comparable with that of light quarks 
after a certain time. 
However, this cannot happen as long as 
$[f_0 (\vpq) I_M(\vpm,t_F)]^{1-r_D(\vpd )} 
\ll f_0(\vpq)/f_{c0}(\vpc)$, 
and we limit our discussion to this case. 
For soft quarks with $f_0 (\vpq) I_M(\vpm,t_F)  \gg  1$, 
we then have
\ber
f^{soft}_D (\vpd,t_F) & \simeq &
f_{c0} (\vpc) \zcdc, \nonumber \\
f^{soft}_\psi (\vpj,t_F)  &\simeq&
\frac {f^2_{c0} (\vpc) r_\psi(\vpj) r_D(\vpd)}
{f_0 (\vpq) \left [ 2 r_D(\vpd)-1 \right ] \nqmc }
{\rm ~for~} r_D  \geq  1, \nonumber \\
&\simeq&
\frac {f^2_{c0} (\vpc) r_\psi(\vpj) r_D(\vpd)}
{f_0 (\vpq) \left [ 2 r_D(\vpd)-1 \right ] \nqmc }
{\rm ~for~} 1>  r_D >  1/2, \nonumber \\
&\simeq&
\frac {f^2_{c0} (\vpc) r_\psi(\vpj) r_D(\vpd)} {f_0 (\vpq) \nqmc } 
\ln  \left [f_0 (\vpq) I_M(\vpm,t_F) \right ]
{\rm ~for~} r_D  = 1/2,  \nonumber \\
&\simeq&
 \frac {f^2_{c0} (\vpc) r_\psi(\vpj) r_D(\vpd)} 
{f_0 (\vpq)  \left [ 1-2 r_D(\vpd) \right ] \nqmc }
\left [f_0 (\vpq) I_M(\vpm,t_F) \right ]^{1-2r_D(\vpd)} \nonumber \\
& & {\rm ~for~} r_D  < 1/2.
\label{charms}
\eer
Note that the above solution of $f^{soft}_\psi (\vpj,t_F)$ 
for $1>  r_D >  1/2$ is valid if $r_D$ is 
not very close to 1/2 so that 
$[f_0(\vpq)/f_{c0}(\vpc)]^{(2r_D-1)/(1-r_D)} \gg 1$ is satisfied, 
and the above solution for $r_D  < 1/2$ is valid if $r_D$ is 
not very close to 1/2 so that 
$[f_0 (\vpq) I_M(\vpm,t_F)]^{1-2r_D} \gg 1$ is satisfied.

\subsection{Effects on elliptic flows at low $\pp$}

Eq.~(\ref{charms}) shows that, contrary to the expectation 
based on the scaling relations for hard quarks, the distribution of soft $D$ 
mesons only depends on the charm quark distribution. 
This is a consequence of the charm quark number conservation, which 
means that (almost) each and every soft charm quark will form a $D$ meson 
via coalescence with a comoving light anti-quark. Since the momentum vector 
of the formed $D$ meson is a constant factor ($1/\zcd$) 
times the momentum of the charm quark, the momentum distribution of 
soft $D$ mesons will be completely determined by the initial 
momentum distribution of the charm quarks. 
Eq.~(\ref{charms}) also leads to
\ber
v^{soft}_{2,D} (\pp)  \simeq  \vtwoc (\zcd \pp ),
\eer 
and this is different from the result for hard $D$ mesons \cite{Lin:2003jy}:
\ber
v^{hard}_{2,D} (\pp)  \simeq  \vtwoc (\zcd \pp )
+\vtwoq (\zqd \pp ). 
\eer

Furthermore, the distributions of soft $J/\psi$ mesons are proportional to 
$f^2_{c0} (\vpc)/f^a_0(\vpq)$ with $a$ being positive
(modulo a logarithmic dependence on $f_0(\vpq)$), and this leads to
\ber
v^{soft}_{2,\psi} (\pp)  \simeq  2 \vtwoc (\pp/\nqm )
-a~\vtwoq (\pp m_q/\nqm/m_c ). 
\eer

For example, if there is no charm elliptic flow, the elliptic flow of 
$J/\psi$ mesons would be negative 
since light quarks have a positive elliptic flow. 
We may understand this correlation from Eq.~(\ref{fc}), 
which shows that the charm quark distribution during the coalescence process 
is inversely correlated to the light quark distribution $f_0(\vpq)$.   
If there is no azimuthal asymmetry in the charm quark momentum distribution 
from the parton phase, $f_c (\vpc,t_0)$ does not depend 
on the azimuthal angle of the charm quark momentum $\vpc$.  
After time $t_0$, however, 
Eq.~(\ref{fc}) shows that the charm quark momentum distribution 
$f_c (\vpc,t)$ will have an azimuthal asymmetry that has an 
opposite sign to the asymmetry in the light quark momentum distribution. 
Because the formation rate of the $J/\psi$ meson distribution 
at any given time during coalescence is proportional to the square of 
the charm meson distribution at that time, 
$J/\psi$ mesons thus have an elliptic flow that has an opposite sign 
as the light quark elliptic flow.

\section{Discussions}

Several simplifying assumptions have been made in this study in order to
obtain analytical results that allow us to see qualitative features
of the effects of quark number conservation. Although we expect some 
key findings of this study to be robust, such as the linear scaling
for low $\pp$ hadrons that dominate the quark number of a flavor 
and the inverse correlation between the soft $J/\psi$ distribution and
the light quark distribution, quantitative predictions are currently
difficult to make. 
Further study is needed to investigate whether the assumptions in
this study are valid for high energy heavy ion collisions and how more 
realistic conditions for quark coalescence affect the results obtained
here. 

To obtain quantitative model predictions, numerical estimates of
quantities such as the coalescence coefficients and the
coalescence momentum scales such as $\ptc$ and $\ptcc$ 
are needed. When the coalescence coefficients are introduced in 
Eqs.~(\ref{coalm1}-\ref{dfbdt}), they are functions of the
hypersurface space-time variable $x$ and the time variable $t_c$ in
general. However, it is possible that they do not depend on these
variables, noting that 
the time integrals of the coalescence coefficients for hard quarks 
are only given by the statistical factor for forming the corresponding
hadron, as shown by the normalization relations of
Eqs.~(\ref{normcm}-\ref{normcb}). 
In order to obtain analytical results, 
in most examples of this study we have assumed that the 
coalescence coefficients have the same time-dependence. 
If the coalescence coefficients do not depend on time, that would just be 
a special case of them having the same time-dependence, therefore all the 
corresponding solutions derived in this study would still be valid. 
On the other hand, if the coalescence 
coefficients have different time-dependencies, soft hadrons that
dominate the quark numbers of each flavor still follow the linear
scaling, while the non-dominant hadrons can have different scaling
relations that depends on how the coalescence coefficients behave as a
function of time, as shown explicitly in Sec.~\ref{sec32}.

We have made the assumption that the
hadronization time scale, $t_F-t_0$, is short compared to the 
time scale for the expansion of the system so that the parton
hypersurface during coalescence is assumed to be unchanged. It
is unclear whether this assumption is valid for high energy heavy ion
collisions. For a first-order phase transition with a large change in
the number of degrees of freedom, the time duration of the mixed
phase can be long. On the other hand, the phase transition of quantum
chromodynamics in equilibrium is a smooth cross-over, and furthermore
the dense matter 
created in heavy ion collisions can be out of equilibrium when
coalescence takes place. In addition, the starting time of the quark
coalescence process corresponds to the time when the effective degrees
of freedom become constituent quarks where gluon degrees of freedom
are no long explicitly present, therefore it is also unknown when this
occurs in the cross-over phase transition. More studies are needed to
incorporate the effect of expansion in this formulation of the quark
coalescence model.

The new formulation of quark coalescence naturally gives the coalescence
probability of quarks and it decreases strongly with increasing
$\pp$. We assume that quarks that do not coalesce will hadronize via
other processes such as fragmentation, and the fragmentation process 
is expected to dominate the particle yields at very high transverse
momentum \cite{Fries:2003vb}. However, particles produced from
fragmentation of partons are not included in the hadron distributions
calculated in this study. Since quark number conservation affects the
scaling relations of soft particles but not hard particles, we do not
expect a significant correction due to neglecting the fragmentation
products.

We note that some of these assumptions are also shared by most studies using 
the previous quark coalescence formulation that does not include quark
number conservation, such as neglecting the internal momentum of partons 
inside a hadron and neglecting the effect from integrating over the full 
hadronization hypersurface. 
Since the parton and hadron distributions are only derived for a local
hypersurface in this study, the effect from integrating over the full
hypersurface is not included. 
Furthermore, the final hadron distributions that we derive 
in this study are the 
distributions just after quark coalescence. Later rescatterings 
in the hadron phase, which could significantly affect hadron momentum 
spectra and particle ratios, have been neglected.
Also, resonance decays have been shown to affect the elliptic flow, 
especially that of pions \cite{Greco:2004ex}; however, we have not
considered multiple hadron species that include resonances.

\section{Conclusions}

We have developed a new formulation of the quark coalescence model 
that includes the quark number conservation. 
We find that scalings between hadron and quark momentum distributions
depend on the quark transverse momentum. For hard quarks, our general
results reproduce the usual scaling relations.  
For softer quarks, however, hadrons that dominate the quark number of  
each flavor exhibit a linear scaling due to the quark number
conservation, while the scalings of non-dominant hadrons depend on
details of the coalescence dynamics. 
For charm mesons we find that, contrary to naive expectations, the
distribution of soft $D$ mesons does not depend on the light quark
distribution while the distribution of soft $J/\psi$ mesons 
is inversely correlated to the light quark distribution. 
Further study is needed to investigate the coalescence dynamics in
order to obtain quantitative predictions. 
These quark-hadron scalings that depend on the particle transverse
momentum can be used to test the quark coalescence model  
as an effective hadronization model for partonic matter, 
and the confirmation would provide new evidence for the formation 
of the quark-gluon plasma in relativistic heavy ion collisions.


\section*{References}
\begin{thebibliography}{99}


\bibitem{Molnar:2003ff}
Molnar D and Voloshin S A
\Journal{\PRL}{}{91}{092301}{2003}

\bibitem{Adler:2003kt}
Adler S S {\it et al}  (PHENIX Collaboration)
\Journal{\PRL}{}{91}{182301}{2003}

\bibitem{Abelev:2007rw}
Abelev B I {\it et al}  (STAR Collaboration)
\Journal{\PRL}{}{99}{112301}{2007}

\bibitem{Lin:2003jy}
Lin Z W and Molnar D
\Journal{\PR}{C}{68}{044901}{2003}

\bibitem{Fries:2003vb}
Fries R J, Muller B, Nonaka C and Bass S A
\Journal{\PRL}{}{90}{202303}{2003}

\bibitem{Greco:2003xt}
Greco V, Ko C M and Levai P
\Journal{\PRL}{}{90}{202302}{2003}

\bibitem{Adler:2003cb}
Adler S S {\it et al}  (PHENIX Collaboration)
\Journal{\PR}{C}{69}{034909}{2004}

\bibitem{Adams:2006ke}
Adams J {\it et al}  (STAR Collaboration)
\Journal{\PRL}{}{98}{062301}{2007}

\bibitem{Zimanyi:ky}
Zimanyi J, Biro T S, Csorgo T and Levai P
\Journal{\PL}{B}{472}{243}{2000}

\bibitem{Yang:2005dg}
Yang C B
\Journal{\JP}{G}{32}{L11}{2006}

\bibitem{Lin:2004en}
Lin Z W, Ko C M, Li B A, Zhang B and Pal S
\Journal{\PR}{C}{72}{064901}{2005}

\bibitem{Zhang:2008zzk}
Zhang B, Chen L W and Ko C M
\Journal{\JP}{G}{35}{065103}{2008}

\bibitem{Dover:1991zn}
Dover C B {\it et al}
\Journal{\PR}{C}{44}{1636}{1991}

\bibitem{Greco:2004ex}
Greco V and Ko C M
\Journal{\PR}{C}{70}{024901}{2004}

\end{thebibliography}
\end{document}